\documentclass[nohyperref]{article}
\usepackage{microtype,graphicx,booktabs,hyperref}
\usepackage[accepted]{icml2022}
\usepackage{amsmath,amssymb,mathtools,amsthm}
\usepackage{multirow,url,comment,caption,subfig,enumitem}
\usepackage[capitalize,noabbrev]{cleveref}

\icmltitlerunning{Zero-shot Blind Image Denoising via Implicit Neural Representations}
\begin{document}
\twocolumn[
\icmltitle{Zero-shot Blind Image Denoising via Implicit Neural Representations}

\begin{icmlauthorlist}
\icmlauthor{Chaewon Kim}{ai}
\icmlauthor{Jaeho Lee}{pos}
\icmlauthor{Jinwoo Shin}{ai,ee}
\end{icmlauthorlist}

\icmlaffiliation{ai}{KAIST AI}
\icmlaffiliation{ee}{KAIST EE}
\icmlaffiliation{pos}{POSTECH EE}

\icmlcorrespondingauthor{Chaewon Kim}{chaewonk@kaist.ac.kr}
\vskip 0.3in
]

\printAffiliationsAndNotice{}

\begin{abstract}
Recent denoising algorithms based on the ``blind-spot'' strategy show impressive blind image denoising performances, without utilizing any external dataset.
While the methods excel in recovering highly contaminated images, we observe that such algorithms are often less effective under a low-noise or real noise regime.
To address this gap, we propose an alternative denoising strategy that leverages the architectural inductive bias of implicit neural representations (INRs), based on our two findings:
(1) INR tends to fit the low-frequency clean image signal faster than the high-frequency noise, and (2) INR layers that are closer to the output play more critical roles in fitting higher-frequency parts. Building on these observations, we propose a denoising algorithm that maximizes the innate denoising capability of INRs by penalizing the growth of deeper layer weights. We show that our method outperforms existing zero-shot denoising methods under an extensive set of low-noise or real-noise scenarios.

\end{abstract}

\section{Introduction}
\label{introduction}

Image denoising is a fundamental yet practical research problem that aims to recover a clean image from a noisy one. Denoising plays an irreplaceable role in tasks where the availability of high-quality images is pivotal for performance, such as classification, object detection, and segmentation;
it is widely known that tiny perturbations on an image can lead to significant performance degradation on downstream tasks \citep{Liu2018denoising, Kaplan2021sensitivity, Michaelis2020benchmarking}. While the problem has been long-explored since the 1950s, image denoisers for a more challenging setup where \textit{no training data or noise information is given} remains remarkably underexplored. Such scenarios are often the case for the most noise-sensitive domains---i.e., denoising is most needed---such as medical and scientific applications where the ground-truth image is expensive to obtain: CT, MRI, live-cell imaging, and fluorescence microscopy~\citep{Weigert18, zhang2019fmd}.

Previous efforts on \textit{``zero-shot'' blind denoising} aim to fill this gap by providing algorithms that denoise on the basis of a single target image. Classic approaches introduce various assumptions on the inherent properties of noise that differs from the original signal and propose to use image filtering~\citep{Dabov07bm3d}, low-rank approximation~\citep{Buades05nlm}, or sparse encoding~\citep{Dong13Sparse} to separate the noise from the image. As these methods critically rely on such assumptions, their performances are typically limited under the setups where the assumptions do not hold. Alternatively, recent work on
deep image prior (DIP; \citet{UlyanovVL17dip}) shows that randomly initialized convolutional neural networks provide a powerful inductive bias about clean image signals, enabling denoising without any explicit assumption on the noise properties. These observations give rise to the following question: Is such convolutional prior the only inductive architectural bias we can use for the images?

In this paper, we explore the potential of the 
architectural inductive bias given by multi-layer perceptrons (MLPs) for blind image denoising. Instead training an MLP that maps a random noise to the image (e.g., as in DIP), we build on the recent line of works on \textit{implicit neural representations} (INR; \citet{sitzmann2019siren}). In this INR approach, a target image is parameterized by a simple MLP that approximates the \textit{image function}, which maps pixel coordinates to the corresponding RGB values. We ask whether this new way of image parameterization also provides a good (or better) inductive prior for image denoising.

\begin{figure*}[ht]
\centering
\includegraphics[width=0.9\textwidth]{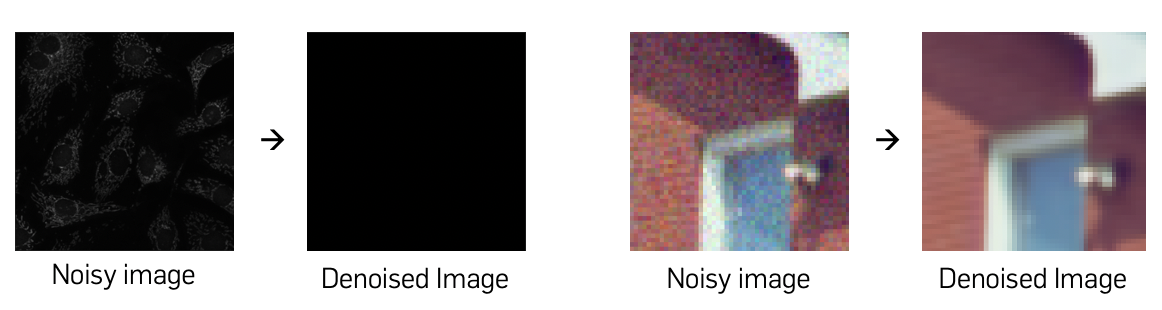}
\caption{Example failure cases of the blind-spot denoising strategy, where we used the state-of-the-art Self2Self \cite{Quan2020s2s} algorithm. \textbf{Left (Real-Noise)}: Real noises are often distributed with a non-zero mean, and blind-spot methods struggle to distinguish the noise from the original signal. \textbf{Right (Low-Noise)}: Blind-spot mechanism performs redundant regularization on the produced image, making the output image too blurry for low-noise cases.  
} \label{fig:s2s}
\end{figure*}

\textbf{Contributions.} We find that implicit neural representations (INR), when trained on a noisy image, have a tendency to fit the underlying clean signal during the early training phase without any external supervision. While the representation eventually fits the noisy image, there exists a window where the generated image looks closer to the clean image than the noisy one. To best utilize this temporal separation of learning noise and signal, we propose a denoising method based on noise-level estimation and layer-selective weight decay. In our experiments on blind noise setups, the proposed INR-based method shows a strong image denoising performance. Especially in the low-noise or real-noise regime, the method outperforms state-of-the-art denoising algorithms based on the ``blind-spot'' strategy \citep{Quan2020s2s}; under these scenarios, interestingly, we observe that the blind-spot denoising algorithms show an unreliable performance (see, e.g., \cref{fig:s2s} for an illustration).

\begin{figure*}[h]
\centering
\subfloat[The inductive architectural bias of INR] {\includegraphics[width=1\columnwidth]{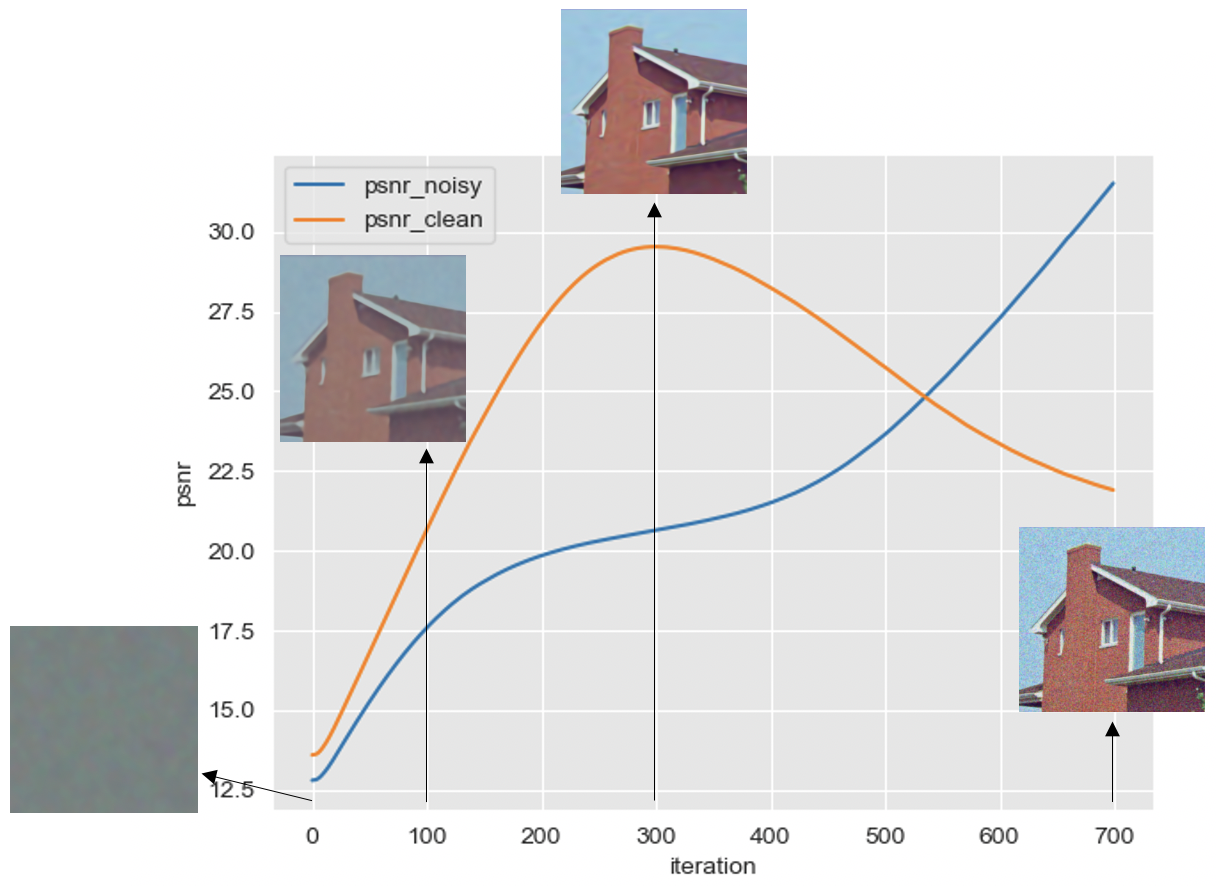}}\subfloat[Signal and Noise Impedances of INR] {\includegraphics[width=0.95\columnwidth]{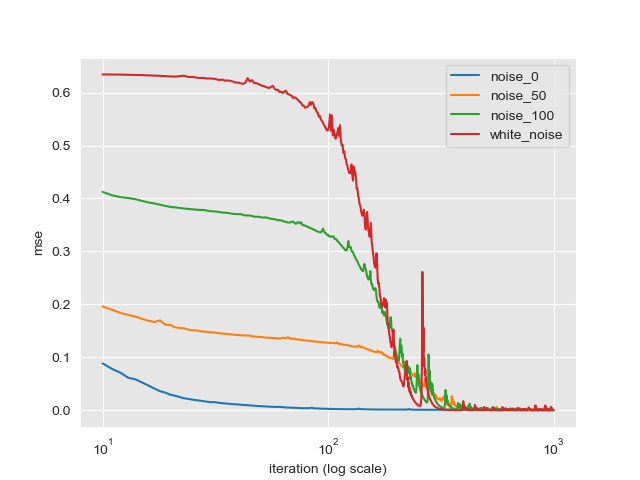}}
\caption{The inductive bias of INR to fit clean image. (a) We fit an INR to an image injected with the Gaussian noise and compare the PSNR of the INR-generated with both the clean image (orange) and the noisy image (blue). INR fits the clean image first, and the noisy image later. (b) We fit INR to a clean image (blue), images contaminated with additive Gaussian noise (orange for $\sigma = 25$ and green for $\sigma=50$), and white noise (red). The learning curve for INR converges faster on a clean image.}
\label{fig:inr_plot}
\end{figure*}

\section{Related Work}\label{sec:related_works}
In this section, we give a brief overview of image denoising algorithms (\cref{ssec:related_denoise}) and existing literature on the inductive bias from neural network architectures (\cref{ssec:related_bias}).

\subsection{Image denoising algorithms} \label{ssec:related_denoise}
Algorithmic approaches to image denoising have a rich history; see, e.g., \citet{BuadesSurvey}. Here, we focus on describing recent methods based on learning.

\textbf{Training with clean-noisy image pairs.} A large number of image denoising algorithms consider a supervised setup, where we use a dataset composed of clean and noisy image pairs of the same scene \citep{yue2019vdn, chen2021hinet, Zamir2021MPRNet}. In the works, neural nets are trained to map noisy images to the corresponding clean image. For example, a popular benchmark method called DnCNN \citep{Zhang2017dncnn} uses deep residual convolutional network for blind denoising. Such supervised methods suffer from three limitations. First, while the methods usually require a large number of training data (due to the high dimensionality of input and output), collecting such clean-noisy image pairs is costly, as it is difficult to gather multiple images with an exact same alignment. Second, training such models usually require a lot of computation. Third, supervisedly trained denoisers tend to perform poorly on domains that differ significantly from the training data; it is known, for instance, that DnCNN \citep{Zhang2017dncnn} trained on datasets of white Gaussian noise performs poorly on real noise datasets such as PolyU \citep{Xu2018polyu}.

\textbf{Training with noisy images.} Another line of research aims to learn denoisers from (unpaired) noisy image datasets. A popular approach is a \textit{blind-spot strategy}, where the goal is to train a model which outputs a noisy image, given the same noisy image as an input. A key idea of the strategy is to avoid learning the identity function by imposing constraints on the model (i.e., blind spots), such as limiting the receptive field that is used to predict each pixel or even dropping input pixels. In \citet{krull2019n2v, Laine2019highqual}, for instance, the receptive field for an output pixel is constrained to not contain the input pixel at the same location. Unfortunately, these methods often fail to perform competitively on real-world images, as such masking procedures may discard useful information. Several recent approaches aim to address this drawback of the blind-spot strategy. R2R \citep{pang2021r2r}, for example, produces two noisy versions of the image by injecting two independent noises into the image and then using them as input/output pairs for training.

\textbf{``Zero-shot'' blind denoising.} Blind denoising considers a more challenging setup where we want to denoise an image without access to auxiliary information of any form (e.g., training data). Previous approaches to this task can be roughly categorized into two: methods relying on the blind-spot strategy, and the methods using a form of inductive bias on images (we discuss the latter in \cref{ssec:related_bias}). Many blind-spot denoising algorithms that are designed to use noisy datasets (e.g., Noise2Self \citep{batson2019n2s}, Noise2Void \citep{krull2019n2v}) can also be used for blind denoising, by using the target image as a training dataset with a single sample. However, their performances are often unsatisfactory---as noted by \citet{Quan2020s2s}---due to its vulnerability to overfitting. To resolve this issue, \citet{Quan2020s2s} propose a refined blind-spot denoising algorithm that utilizes a dropout-based ensemble designed exclusively for single-image denoising. Still, a shortfall remains; the blind-spot strategy enforces the same degree of `smoothing' on all different types and powers of noises, often resulting in over-smoothing on real-noise or low-noise settings (as illustrated in \cref{fig:s2s}).

\subsection{Inductive bias from neural net architecture} \label{ssec:related_bias}
The term \textit{inductive bias} denotes the bias of a learning procedure to prefer a solution over another, independent of the observed data \citep{Mitchell1980bias}. Such bias can be beneficial when it helps search for solutions that generalize on unseen data. One way to introduce the inductive bias purposefully is to use different model architecture. For example, convolutional networks are known to introduce beneficial bias for visual recognition by encouraging the predictor to utilize the local structures of an image \citep{zeiler2014}. In contrast, fully-connected networks have been long considered less effective for visual tasks.

\textbf{Deep image priors.} The inductive bias of convolutional architecture has been actively utilized in the context of image denoising. One of the most popular approaches, based on \textit{deep image priors} (DIP), has been introduced by \citet{UlyanovVL17dip}. In the work, the authors propose to train a fully convolutional model which predicts the target noisy image given the random noise as an input. While the model eventually learns to generate the noisy image, it was empirically observed that the model tends to generate cleaner version of the image in the middle of training, which can be captured by early stopping. \citet{Chakrabarty2019Spectralbias} provides an extensive analysis of this early stopping step and points out the similarity with the low-pass filtering. In this work, we show that fully-connected networks---as opposed to convolutional ones---also possess a beneficial bias for image denoising.

\section{The Bias of Implicit Neural Representations}\label{sec:method}

In this section, we study the inductive bias of implicit neural representations and describe our algorithm which aims to maximize and utilize the bias to denoise the target image without any auxiliary information.

\begin{figure*}[ht]
    \centering
    \subfloat[INR trained until convergence]{
    	\label{subfig:correct}
    	\includegraphics[width=0.9\textwidth]{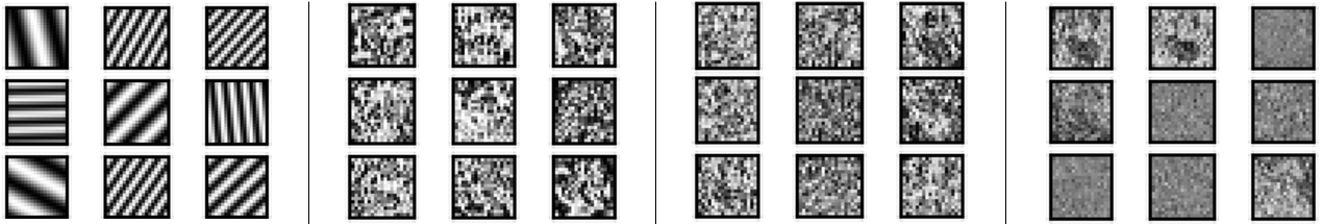}} 
    	
    \subfloat[INR trained to the optimal early stopping epoch]{
    	\label{subfig:notwhitelight}
    	\includegraphics[width=0.9\textwidth]{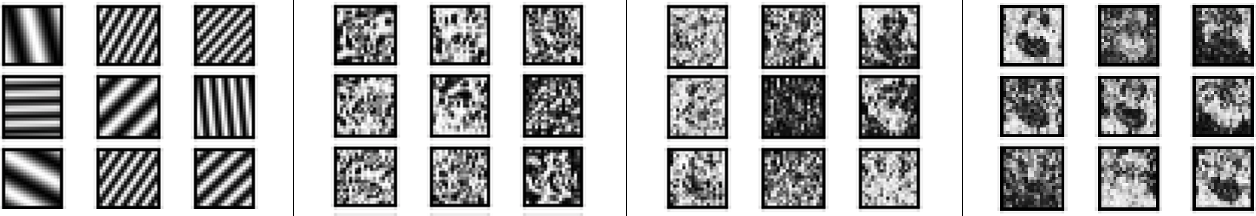}} 
    \caption{Visualization of implicit features of INRs fitted to a noisy image. We used an astronaut image from the \textit{skimage} library, corrupted with white additive Gaussian noise ($\sigma=25$). Each box represents the activation of a neuron belonging to the target hidden layer. We randomly selected 9 neurons from each of four hidden layers in INR, with the leftmost column being the first hidden layer that is closest to the input layer, and the rightmost column being the last hidden layer that is closest to the output layer. We observe that (1) early layers capture the global coarse features while deeper layers match local fine details, and (2) early stopping prevents the model from fitting the high-frequency noise.}
    \label{fig:implicit_features}
\end{figure*}

\subsection{Preliminaries: implicit neural representation}

Before we describe its bias, we briefly introduce the notion of implicit neural representations (INRs).
Implicit neural representation is a form of data representation, that aims to parameterize the coordinate-to-value mapping underlying the target data using MLPs---see, e.g., \citet{Stanley2007Compositional,sitzmann2019siren}. More formally, consider the case of RGB image data. The target function to be learned may be represented by a function $g: \mathbb{R}^{2} \to \mathbb{R}^3$, where the input dimensions correspond to $(x,y)$ coordinates of each pixel and the output dimensions correspond to $(R, G, B)$ channels of the pixels. We train an MLP $f_{\theta}$ (parameterized by network weights $\theta$) with two input neurons and three output neurons to approximate this target function $g$, using the squared loss
\begin{align}
    \mathcal{L}(\theta) = \sum_{c \in \mathcal{C}}\|f_{\theta}(c) - g(c)\|_2^2, \label{eq:self_supervised_loss}
\end{align}
where $\mathcal{C}$ denotes the set of all pixel coordinate pairs $c = (x,y) \in \mathbb{R}^2$ in the target image; for an image with $256 \times 256$ pixels, for instance, the coordinate pair set may be the grid
\begin{align}
\mathcal{C} = \left\{0,\frac{1}{255},\frac{2}{255},\ldots,\frac{255}{255}\right\}_,^2
\end{align}
in which each pair denotes the center coordinate of the pixel.

The MLP constituting the INR can be constructed in many ways; \citet{tancik2020fourfeat} proposes to use ReLU neural network with a Fourier feature encoding in the first layer, while \citet{sitzmann2019siren} proposes a model named SIREN which uses sinusoidal activation functions in every layer (except the last layer). Throughout this paper, unless noted otherwise, we use SIREN as our base INR with a uniform number of hidden neurons in every layer.

\subsection{The inductive bias of implicit neural representations}\label{ssec:bias}

When trained on a noisy image until convergence, INR eventually fits the noisy image with only a very small distortion. During the training, however, we find that INR tends to produce an image that is closer to the clean image than the noisy one, without any explicit supervision on the loss.

We give an illustrative example in \cref{fig:inr_plot}(a);
we train an INR to fit a noisy image and compare the PSNR of the INR-generated image with both clean and noisy versions of the image. We observe that the clean image PSNR achieves its peak performance during the early phase ($\approx$ 300 iterations) and then decreases. On the other hand, the noisy image PSNR increases steadily but relatively slower than the clean image PSNR, not exceeding the clean image PSNR before 500 iterations. Furthermore, interestingly, we find that there is a short ``dragging'' period in the continual growth of the noisy image PSNR, which \textit{coincides} with the peak of clean image PSNR.

These observations suggest that fitting an INR to a noisy image can be divided into roughly two temporal phases. (1) Clean image fitting: INR first fits the clean image, which is relatively easier (in some sense) for the INR to express. (2) Noise fitting: Fine modifications to the INR weights are made, fitting the hard-to-fit noise component. Given this hypothesis, our goal is to design a denoising method that maximizes this signal-noise separating property of INR, and then utilizes the separation to denoise the image.

We also look at the ``inertia'' of INR architecture in learning signal and noise in \cref{fig:inr_plot}(b). We followed the analysis of \citet{UlyanovVL17dip} to capture the inertia—which they denote by the term impedance—via measuring the MSE loss trajectory while fitting a clean image, a clean image contaminated by additive noise, and the white noise.  First, INR fits the clean image faster with a steep slope in the early training than noisy images. INR also eventually fits the noise, converging at near-zero MSE. We analyze that the effectiveness of INR on denoising comes from its ability to learn the clean image fast.


\subsection{Proposed method}\label{ssec:method}

Given the signal-noise separating property of INR, a straightforward method for denoising an image is an early stopping, i.e., stop fitting INR after training for some number of steps. However, as we are considering the blind zero-shot denoising setup where we have no access to any auxiliary information, it is hard to know \textit{when} to stop training.

\begin{table*}[t]
\caption{Quantitative comparison (mean PSNR(dB)) of different methods for low-noise denoising. We draw $5$ different noise levels ($\sigma, \alpha$) and report the average PSNR over the values; these noise levels are shared overall methods for a fair comparison. Bold denotes the highest PSNR overall methods.}
\label{tab:lownoise}
\centering
\scalebox{1.0}{
\begin{tabular}{c|c|cccccc}
\toprule
Noise & Image & DIP & N2V & N2S & N2F & S2S & \textbf{INR(Ours)}\\
\midrule
\multirow{9}{*}{\parbox{4cm}{Gaussian Noise\\ $\sigma \in [0,25]$}} & House & 33.51 & 33.72 & 34.82 & 33.88 & \textbf{34.98} & 33.94\\
& F16 & 35.75 & 34.73 & 35.32 & 35.71 & 35.86 & \textbf{37.35} \\
& Peppers & 31.32 & 30.82 & 31.37 & 31.42 & 31.77 & \textbf{31.78}\\
& Baboon & 28.76 & 25.38 & 26.83 & 26.86 & 27.27 & \textbf{28.76}\\
& Lena & 34.36 & 33.24 & 33.87 & 34.36 & 34.26 & \textbf{36.36}\\
& Kodim01 & 29.83 & 27.39 & 29.02 & 30.31 & 33.47 & \textbf{35.16}\\
& Kodim02 & 33.38 & 32.84 & 32.92 & 33.40 & 35.42 & \textbf{35.89}\\
& Kodim03 & 34.75 & 33.06 & 34.21 & 36.10 & 37.89 & \textbf{39.22} \\
& Kodim12 & 35.27 & 33.84 & 33.89 & 34.98 & 35.31 & \textbf{35.88} \\
\midrule
\multirow{9}{*}{\parbox{4cm}{Poisson-Gaussian Noise\\ $\alpha \in [50,100],  \sigma \in [0,25]$}} & House & 32.17 & 31.20 & 31.00 & 31.73 & 32.18 & \textbf{32.30}\\
& F16 & 34.65 & 32.02 & 33.84 & 34.53 & 34.62 & \textbf{34.77} \\
& Peppers & \textbf{32.64} & 30.37 & 31.28 & 32.21 & 31.99 & 32.38 \\
& Baboon & 24.51 & 22.10 & 22.81 & 23.39 & 23.45 & \textbf{24.97}\\
& Lena & 25.04 & 23.19 & 23.87 & 25.01 & 24.95 &\textbf{25.87} \\
& Kodim01 & 25.23 & 24.94 & 25.22 & 25.12 & 26.13 & \textbf{26.82} \\
& Kodim02 & 28.31 & 26.30 & 26.38 & 27.17 & 28.37 & \textbf{28.64} \\
& Kodim03 & 25.78 & 24.39 & 24.90 & 25.01 & 25.76 & \textbf{25.96} \\
& Kodim12 & 21.80 & 20.05 & 20.36 & 21.09 & 21.71 & \textbf{21.88} \\
\bottomrule
\end{tabular}
}
\end{table*}

We propose to use the following simple early stopping criterion: We stop training whenever the mean-squared error (MSE) between the INR-generated image and the target noisy image gets smaller than the noise level estimated from the target image. In other words, when $\hat{n}(g)$ denotes the noise power estimate of the image function $g$, we stop training when
\begin{align}
    \sum_{c\in\mathcal{C}}\|f_{\theta}(c) - g(c)\|^2_2 \le \hat{n}^2(g). \label{eq:stopping}
\end{align}
This stopping criterion is designed considering the ideal case, where the INR bias provides a perfect temporal separation of the image and noise fitting (i.e., the underlying clean signal is fit before fitting the noise). Indeed, if the INR-generated image is perfectly identical to the clean image, the stopping criterion (\cref{eq:stopping}) will be satisfied with equality.

For the noise estimation, we find that the nonparametric method of \citet{Chen2015noiseestimation} fits well for our purpose of blind denoising, as it imposes a minimal set of assumptions. We note, however, that our method can be combined with other noise estimation methods as well whenever we can utilize further auxiliary information about the underlying noise-generating procedure.

\textbf{Selective Weight Decay.} In addition, to strengthen the INR bias, we propose to apply weight decay selectively on the \textit{last two layers} of INR. The main motivation comes from our observation that the latter INR layers contribute more to fitting the noise signal than earlier layers. More concretely, \cref{fig:implicit_features} gives a visualization of individual neuron outputs in each layer, of an INR with four hidden layers trained on a noisy image. 
From the figure, we observe a hierarchy of features; earlier layer neurons mainly represent low-frequency patterns while latter layer neurons generate fine-grained details. One notable difference between the fully-trained and early-stopped model is the behavior of latter-layer neurons.
In particular, comparing with INR at the optimal early stopping epoch, INR trained until convergence differs mainly in their latter layer neuron outputs; many of latter layer outputs lose semantic information and converges to a noise. On the other hand, earlier layer outputs barely change. We note that a similar finding has been reported by \citet{chen2021featurematching}, where the authors use the smooth interpolation of implicit features for 3D shape deformation.

\begin{table*}[t]
\caption{Denoising performance on FMD dataset, measure in mean PSNR (left) and SSIM (right). The noise level is controlled via the number of images averaged during the data-generating process. Bold denotes the best performance.}\label{tab:fmd}
\centering
\begin{tabular}{@{}lcccccc@{}}
\toprule
& \multicolumn{5}{c}{Number of images averaged $(S)$}\\
\cline{2-6}
 & 1 & 2 & 4 & 8 & 16 \\
\toprule
DIP & 31.20 / 0.698 & 33.98 / 0.822 & 35.05 / 0.882 & 38.20 / 0.732 & 39.34 / 0.839 \\
N2V & 28.15 / 0.590 & 29.87 / 0.728 & 33.42 / 0.804 & 38.18 / 0.728 & 37.80 / 0.734 \\
N2S & 29.34 / 0.594 & 29.19 / 0.690 & 30.28 / 0.638 & 36.12 / 0.630 & 38.28 / 0.784 \\
S2S & 30.71 / 0.595 & 30.31 / 0.802 & 26.69 / 0.470 & 39.21 / 0.715 & 36.00 / 0.550\\
INR & \textbf{34.01 / 0.809} & \textbf{36.83 / 0.863} & \textbf{39.76 / 0.903} & \textbf{40.01 / 0.968}  & \textbf{41.56 / 0.977} \\
\bottomrule
\end{tabular}
\end{table*}
\begin{table}[ht]
\caption{Denoising performance on PolyU dataset, measured in mean PSNR overall images. For N2V and N2S, we use $70$ randomly selected noisy images for training and the remaining $30$ images for testing. Bold denotes the best performance.}\label{tab:polyu}
\centering
\scalebox{0.75}{\begin{tabular}{@{}lcccccccc@{}}
\toprule
Method & cBM3D & DIP & N2V & N2S & cDnCNN & S2S & \textbf{INR} \\
\midrule
PSNR & 36.98 & 36.95 & 34.08 & 35.46 & 37.55 & 37.52 & \textbf{37.71} \\ \bottomrule
\end{tabular}}
\end{table}

\section{Experiment}\label{sec:exp}

In this section, we validate the empirical performance of the INR-based denoiser proposed in \cref{ssec:method} under the setting of \textit{zero-shot blind image denoising}, where the denoiser is given only a single image to denoise, without any auxiliary information.

\subsection{Setup}

We focus on the low-noise and real-noise regime, where the blind spot denoising strategy often fails to achieve reasonable performance (as described in \cref{fig:s2s}).

\textbf{Low-noise dataset.} For the low-noise setup, we contaminate clean images from the Set9 dataset \citep{Dabov07bm3d} with either Gaussian noise or Poisson-Gaussian noise (see, e.g., \citet{foi08poissongaussian,zhang2019fmd}) with small random variance. More specifically, for the Gaussian noise, we contaminate each pixel value as
\begin{align}
    s_{\text{noisy}} = s_{\text{clean}} + \zeta_g, \quad \zeta_g &\sim \mathcal{N}(0,\sigma^2),\nonumber \\
    \sigma &\sim \mathcal{U}([0,25]), \label{eq:gauss_noise}
\end{align}
where $s_{\text{noisy}},s_{\text{clean}}$ denotes the noisy and clean signals, respectively, and $\mathcal{N}, \mathcal{U}$ denotes Gaussian and Uniform distributions, respectively. The noise level $\sigma$ is drawn once for each image and shared for all pixels in the image. For the Poisson-Gaussian noise, we contaminate the image as
\begin{align}
    s_{\text{noisy}} = s_{\text{clean}} + \zeta_g + \zeta_p,
\end{align}
where $\zeta_g$ is defined as in \cref{eq:gauss_noise} and $\zeta_p$ is distributed dependently on $s_{\text{clean}}$ in a way that
\begin{align}
    \frac{s_{\text{clean}} + \zeta_p}{\alpha} \sim \mathcal{P}\left(\frac{s_{\text{clean}}}{\alpha}\right)
\end{align}
holds, with $\mathcal{P}$ being the Poisson distribution and $\alpha$ drawn uniformly randomly from $[50,100]$. Similarly, $\alpha$ is drawn once and shared for all pixels in an image.

\textbf{Real-noise dataset.} For the real-noise setup, we use Fluorescence Microscopy Denoising (FMD) dataset \citep{zhang2019fmd} and PolyU dataset \citep{Xu2018polyu}. FMD dataset covers the three most widely used real-noise microscopy images: confocal, two-photon, and wide-field. To construct this dataset, high-noise samples were first collected by using the lowest excitation power for the microscope laser/lamp, and low-noise samples of different degrees are generated by averaging multiple ($S = 2^0,2^1,\ldots,2^4$) high-noise samples with the same view. Ground-truth samples are generated by averaging $50$ samples. PolyU dataset is a real-world image dataset with a similar data collection procedure.

\textbf{Baselines.} We compare our method mainly against existing blind image denoising techniques, e.g., Self2Self (S2S; \citet{Quan2020s2s}), Noise2Fast (N2F; \citet{Lequyer2021}) and deep image prior (DIP; \citet{UlyanovVL17dip}). For DIP, we use the standard encoder-decoder model with 1D convolutional layers, with 256 channels and 10 layers. We also compare against the single-sample variants of Noise2Void (N2V; \citet{krull2019n2v}) and Noise2Self (N2S; \citet{batson2019n2s}). For the PolyU dataset, we follow the evaluation procedure of \citet{Quan2020s2s} to compare with cBM3D \citep{Dabov07bm3d}, cDnCNN \citep{Zhang2017dncnn}, N2S, and N2V that are trained with $70$ training images; for these methods, we evaluate over the remaining $30$ images.

\textbf{Model and hyperparameters.} We use SIREN \citep{sitzmann2019siren} architecture with $6$ hidden layers. The width of the INR is selected proportionally to the resolution of the image; for an image with height $H$ and width $W$, we used the width $256 \times \frac{H \times W}{512 \times 512}$. We use the same weight decay hyperparameter $\lambda = 0.001$ throughout all experiments.

\subsection{Result} \label{ssec:exp_main}
The experimental results for the low-noise setup on the Set9 images are given in \cref{tab:lownoise}. For both Gaussian and Poisson-Gaussian noises, we observe that the INR consistently achieves the best (or comparable) performance.

The results for the real-noise setup are given in \cref{tab:fmd} and \cref{tab:polyu} for FMD and PolyU dataset, respectively. For all degrees of noise levels, INR-based denoising achieves the best performance among all methods. We note, interestingly, that the performance of S2S is not aligned with the noise level. In particular, the performance of S2S is especially low at $S=4$, while showing better performance at $S=2, 8$. In contrast, performances of the methods not adopting the blind-spot strategy (DIP, INR) are well-aligned with the noise intensity.

\section{Conclusion}\label{sec:conc}

In this paper, we presented a blind denoising technique that uses the architectural inductive bias of implicit neural representation (INR). Without any explicit noise modeling or training, the INR-based denoiser first learns low-frequency features within a single image. We empirically proved that this bias enables INR to be a more appropriate architecture for inverse problems compared to convolutional encoder-decoder structures. We further found that deeper layers are most responsible for learning local high-frequency details. With this observation, we penalized deeper layers by applying weight decay regularization to improve the denoising performance. Extensive experiments, especially in a setting where the state-of-the-art blind-spot methods fail to denoise, showed that INR outperforms other zero-shot-based methods. This inductive bias and continuous nature of INR can be leveraged for other image restoration tasks, which we leave for further investigation.

\bibliography{contents/reference}
\bibliographystyle{icml2022}

\end{document}